\documentclass[seceq]{ptptex}

\usepackage{amsmath}
\usepackage{bm}
\usepackage[dvips]{graphicx}
\newcommand{\be}{\begin{equation}}

\newcommand{\ee}{\end{equation}}
\newcommand{\bea}{\begin{eqnarray}}
\newcommand{\eea}{\end{eqnarray}}

\title{
Phase diagram and pion modes of the neutral Nambu-Jona Lasinio model with the Polyakov loop%
}

\author{
Marco \textsc{Ruggieri}%
}

\inst{
I.N.F.N., Sezione di Bari, I-70126 Bari, Italy,\\Universit\`a di Bari, I-70126 Bari, Italy
 }

\abst{I discuss recent results obtained by our group within the Nambu-Jona Lasinio model with the Polyakov loop. I
focus on the two flavor version of the model, discussing the phase diagram in the $\mu-T$ plane in the case in which
electrical neutrality is required. One of the results is that pion condensation is not observed in our model when a
physical mass for the light quarks is considered.}

\begin{document}

\maketitle

\section{Introduction}
The study of strong interactions by means of effective models, which are simpler to manage than Quantum Chromodynamics
(QCD), is nowadays very popular. Among the various effective models of QCD the Nambu-Jona Lasinio model (NJL) is
widely used since it allows for a simple and non perturbative discussion of chiral symmetry breaking and related
phenomena~\cite{Nambu:1961tp,revNJL}. The NJL model can be improved by the introduction of a background temporal static
gluon field, coupled to quarks via the QCD covariant derivative. The gluon field is related to the expectation value of
Polyakov loop $\Phi$~\cite{Polyakovetal}, and the new model is called Polyakov-NJL (PNJL in the
following)~\cite{Meisinger:1995ih,Fukushima:2003fw,Ratti:2005jh,Roessner:2006xn,Ghosh:2007wy,
Kashiwa:2007hw,Schaefer:2007pw,Ratti:2007jf,Sasaki:2006ww,Megias:2006bn,Zhang:2006gu,Ciminale:2007ei,
Abuki:2008tx,Fu:2007xc,Ciminale:2007sr}

In the PNJL model an effective potential for $\Phi$ is added by hand to the quark lagrangian, and $\Phi$ is coupled to
the quarks via the QCD covariant derivative. The value of $\langle\Phi\rangle$ in the ground state, as well as other
quantities of interest like the chiral and/or the pion condensate, are obtained by minimization of the thermodynamic
potential.

In this proceeding we summarize some of our recent results presented in Ref.~\cite{Abuki:2008tx}, where we have studied
the phase diagram and the pion modes of the electrically neutral two flavor PNJL model. Our study has some overlap with
Refs.~\cite{Zhang:2006gu,Ebert:2005wr}

One of the results of our work is that when electrical neutrality is required, pion do not condense in the ground
state: as explained in our original paper~\cite{Abuki:2008tx} this is mainly due to the fact that we use quarks with a
finite current mass. Moreover we investigate on the pions and $\sigma$ mass spectra. We find that the qualitative
behavior of the masses resembles that obtained in the NJL model. We close this paper by studying the possibility that a
bound state with the quantum numbers of the pions can be formed above the chiral phase transition.

\section{Formalism and results\label{Sec:for}}
The Lagrangian density of the two flavor PNJL model considered here is given by~\cite{Fukushima:2003fw,Abuki:2008tx}
\begin{equation}
{\cal L}=\bar{e}(i\gamma_\mu\partial^\mu + \mu_e \gamma_0)e +  \bar\psi\left(i\gamma_\mu D^\mu + \hat\mu\gamma_0
-m\right)\psi + G\left[\left(\bar\psi \psi\right)^2 + \left(\bar\psi i \gamma_5 \vec\tau \psi\right)^2\right] - {\cal
U}[\Phi,\bar\Phi,T]~, \label{eq:Lagr}
\end{equation}
In the above equation $e$ denotes the electron field;  $\psi$ is the quark spinor with Dirac, color and flavor indices
(implicitly summed). $m$ corresponds to the bare quark mass matrix; we assume from the very beginning $m_u = m_d$. The
covariant derivative is defined as usual as $D_\mu =
\partial_\mu -i A_\mu$. The gluon background field
$A_\mu=\delta_{0\mu}A_0$ is supposed to be homogeneous and static, with $A_0 = g A_0^a T_a$ and $T_a$, $a=1,\dots,8$
being the $SU(3)$ color generators with the normalization condition $\text{Tr}[T_a,T_b]=\delta_{ab}$. $\vec{\tau}$ is a
vector of Pauli matrices in flavor space. $\mu$ is the chemical mean quark chemical potential, related to the conserved
baryon number; $\mu_e = - \mu_Q$ and the quark chemical potential matrix $\hat\mu$ is defined in flavor-color space as
\begin{equation}
\hat\mu=\left(\begin{array}{cc}
  \mu-\frac{2}{3}\mu_e & 0 \\
  0 & \mu + \frac{1}{3}\mu_e \\
\end{array}\right)\otimes\bm{1}_c~,\label{eq:chemPot}
\end{equation}
where $\bm{1}_c$ denotes identity matrix in color space.

In Eq.~\eqref{eq:Lagr} $\Phi$, $\bar\Phi$ correspond to the normalized traced Polyakov loop $L$ and its hermitian
conjugate respectively. The term ${\cal U}[\Phi,\bar\Phi,T]$ is the effective potential for $\Phi$, $\bar\Phi$. Several
forms of this potential have been suggested in the literature, see for
example~\cite{Fukushima:2003fw,Ratti:2005jh,Roessner:2006xn,Ghosh:2007wy}. In this paper we adopt the following
logarithmic form~\cite{Roessner:2006xn},
\begin{equation}
\frac{{\cal U}[\Phi,\bar\Phi,T]}{T^4} = -\frac{b_2(T)}{2}\bar\Phi\Phi + b(T)\log\left[1-6\bar\Phi\Phi + 4(\bar\Phi^3 +
\Phi^3) -3(\bar\Phi\Phi)^2\right]~,\label{eq:Poly}
\end{equation}
with
\begin{equation}
b_2(T) = a_0 + a_1 \left(\frac{\bar T_0}{T}\right) + a_2 \left(\frac{\bar T_0}{T}\right)^2~,~~~~~b(T) =
b_3\left(\frac{\bar T_0}{T}\right)^3~.\label{eq:lp}
\end{equation}
The reader is referred to Ref.~\cite{Roessner:2006xn} for numerical values of the coefficients.

We work in the mean field approximation. In order to study chiral symmetry breaking and to allow for pion condensation
we assume that in the ground state the expectation values for the following operators may
develop~\cite{Zhang:2006gu,Abuki:2008tx,Ebert:2005wr},
\begin{equation}
\sigma = \left<\bar\psi \psi\right>~,~~~~~\pi = \left<\bar\psi i \gamma_5 \tau_1 \psi\right>~.\label{eq:condensates}
\end{equation}
In the above equation a summation over flavor and color is understood.  The thermodynamical potential $\Omega$ can be
obtained by integration over the fermion fields in the partition function of the model~\cite{Abuki:2008tx}
\begin{eqnarray}
\Omega &=& -\left(\frac{\mu_e^4}{12\pi^2} + \frac{\mu_e^2 T^2}{6} + \frac{7\pi^2 T^4}{180}\right) + {\cal
U}[\Phi,\bar\Phi,T] + G\left[\sigma^2 + \pi^2\right] \nonumber\\
&&~~~~~- T\sum_n\int_0^\Lambda\frac{d^3{\bm p}}{(2\pi)^3}~\text{Tr}~\text{log}\frac{S^{-1}(i\omega_n,{\bm p})}{T}~,
\end{eqnarray}
where the sum is over fermion Matsubara frequencies $\omega_n = \pi T(2n+1)$, and the trace is over Dirac, flavor and
color indices. $\Lambda$ is an ultraviolet cutoff to ensure convergence of the momentum integral. The inverse quark
propagator $S^{-1}$ is easily derived by Eq.~\eqref{eq:Lagr} (the interested reader can find the calculational details
in our paper~\cite{Abuki:2008tx}). The ground state of the model is defined by the values of $\sigma$, $\pi$, $\Phi$,
$\bar\Phi$ that minimize $\Omega$ and that have a vanishing total charge; the latter condition is equivalent to the
requirement
\begin{equation}
\frac{\partial\Omega}{\partial\mu_e}=0~.
\end{equation}
The parameters $m$, $G$ and $\Lambda$ are given by~\cite{Roessner:2006xn}
\begin{equation}
m=5.5~\text{MeV}~,~~~~~G=5.04~\text{GeV}^{-2}~,~~~~~\Lambda=650.9~\text{MeV}~,
\end{equation}
which fix, at zero temperature and zero chemical potential, the pion mass $m_\pi=139.3$ MeV, the pion decay constant
$f_\pi=92.3$ MeV and the chiral condensate $\langle\bar u u\rangle = -(251~\text{MeV})^3$. Finally we chose $\bar{T}_0
= 208$ MeV in the Polyakov loop effective potential~\cite{Schaefer:2007pw}.

\begin{figure}[t!]
\begin{center}
\includegraphics[width=8cm]{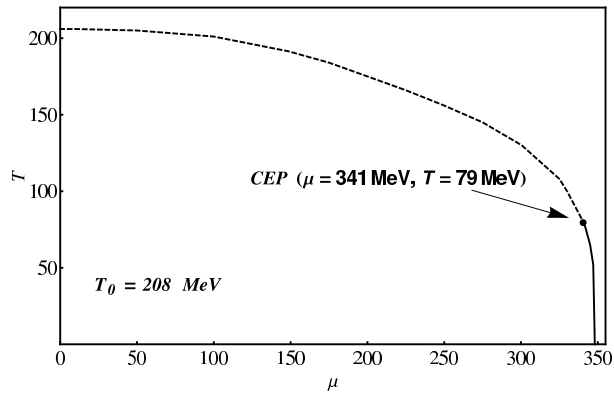}
\end{center}
\caption{\label{FIG:pd}Phase diagram of the electrically neutral two flavor PNJL model. Dashed line corresponds to the
chiral crossover; solid line describes the first order chiral transition. At $\mu=0$ the critical point is located at
$T=206$ MeV; at $T=0$ the chiral transition is found at $\mu=348$ MeV. The black dot denotes the critical end point
(CEP), located at $(\mu_E,T_E) = (342,79)$ MeV.}
\end{figure}

In Fig.~\ref{FIG:pd} we plot the phase diagram of the electrically neutral two flavor PNJL model. Dashed line
corresponds to the chiral crossover; solid line describes the first order chiral transition. At each value of $\mu$ the
crossover is identified with the inflection point of the chiral condensate. Analogously the first order transition is
defined by the discontinuity of $\sigma$. At $\mu=0$ the chiral crossover occurs at $T=206$ MeV. For comparison, the
inflection point of the Polyakov loop (which is commonly associated to the deconfinement crossover) at $\mu=0$ is
located at $T=180$ MeV. This implies that in the model under consideration the two crossovers are mismatched by a
temperature of the order of 10 MeV. Moreover, at $T=0$ the chiral phase transition is of first order and is found at
$\mu=348$ MeV. The black dot denotes the critical end point (CEP), located at $(\mu_E,T_E) = (342,79)$ MeV.

As explained in detail in our original paper~\cite{Abuki:2008tx}, we find a vanishing pion condensate for each value of
$\mu$ and $T$ once electrical neutrality is required. This result seems in contradiction with the results obtained by
Ebert and Klimenko in Ref.~\cite{Ebert:2005wr}, where they show that at $T=0$ and in the neutral phase a narrow window
in $\mu$ there exists where pions condense. The reason of the discrepancy is easily understood by observing that Ebert
and Klimenko work in the chiral limit, thus putting $m=0$ in the quark lagrangian. On the other hand we consider
massive quarks. The value $m=5.5$ MeV used by us to reproduce the vacuum pion mass favors the vacuum alignment to the
chiral condensed phase with $\sigma\neq0,~\pi=0$~\cite{Abuki:2008tx,forthcoming}, in the same way as an external
magnetic field ${\bm H}$ along a certain direction in space favors the alignment of the spins in a Weiss domain along
${\bm H}$ itself. This is not surprising at all: as a matter of comparison, at $\mu=0$ the condition $\mu_I > m_\pi$
must be fulfilled in order to observe pion condensation, with $\mu_I$ the isospin chemical potential. In the chiral
limit $m_\pi=0$ thus an infinitesimally small value of $\mu_I$ is enough to favor pion condensation. On the other hand
a light current quark mass $m\approx5$ MeV implies $m_\pi\approx 140$ MeV, thus a $\mu_I \geq 140$ is needed to observe
pion condensation. The comparison between our results and those obtained by Ebert and Klimenko suggests that the pion
condensate is very sensitive to the value of the current quark mass.    A detailed study of the evolution of the ground
state of strongly interacting quark matter as $m_0$ is increased from zero to its physical value, at zero and finite
temperature and/or chemical potential, will be the subject of a forthcoming paper~\cite{forthcoming}.

Next we turn  to the computation of the masses of the pion modes in the two flavor and electrically neutral PNJL model.
The $\sigma$ mode is not considered here for obvious space limitations; for the same reason we skip all the
calculational details, the interested reader is referred to our original paper~\cite{Abuki:2008tx}. The pion mass is
defined as pole of the pion propagator at rest, the latter being the solution of the Schwinger-Dyson equation for the
pseudoscalar mesonic correlator. Its derivation in the context of the NJL model is in Ref.~\cite{revNJL}. The same
equation is valid in the PNJL model, see for example the clear derivation in~\cite{Hansen:2006ee}.

\begin{figure}[t!]
\begin{center}
\includegraphics[width=7cm]{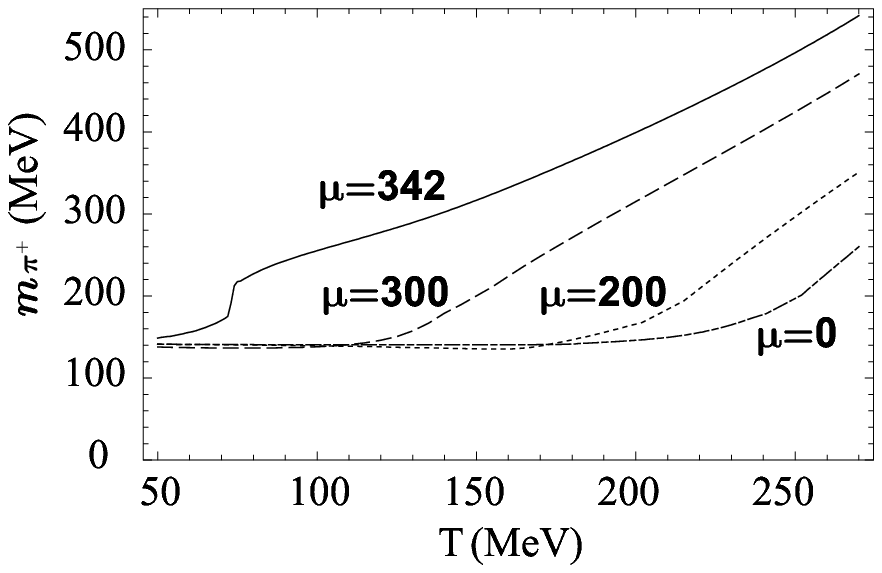}~~~~~\includegraphics[width=7cm]{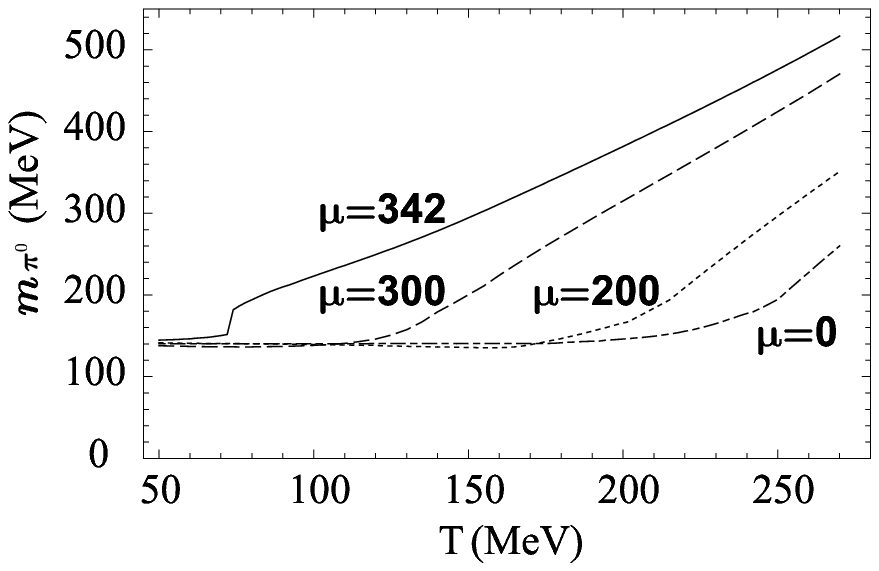}
\end{center}
\caption{\label{FIG:pp}Masses of the charged pions (left panel) and of the neutral pion (right panel) in the
electrically neutral phase, as a function of the temperature, for different values of the quark chemical potential.}
\end{figure}

In Fig.~\ref{FIG:pp} we plot the masses of the charged pions (left panel) and of the neutral pion (right panel) in the
electrically neutral phase, as a function of the temperature, for different values of the quark chemical potential. The
behavior of the pseudoscalar modes as the temperature is increased is qualitatively the same observed in the NJL
model~\cite{revNJL}.

\begin{figure}[t!]
\begin{center}
\includegraphics[width=10cm]{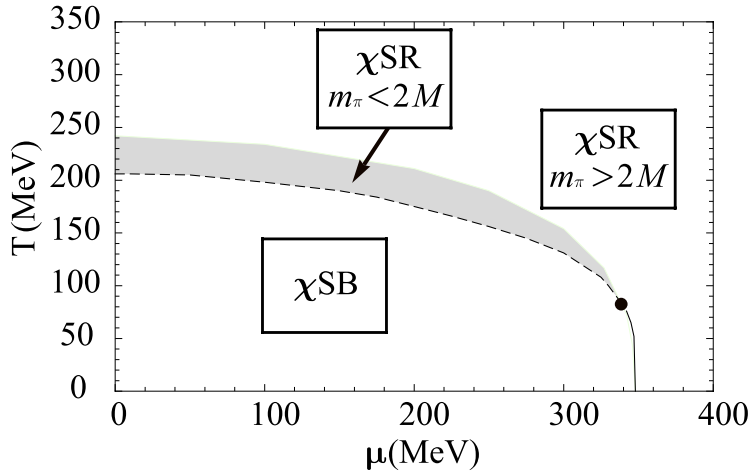}
\end{center}
\caption{\label{FIG:ccc}Region of existence of bound states in the $\mu-T$ plane. $\chi SB$ denotes the region where
chiral symmetry is broken; $\chi SR$ denotes the region where chiral symmetry is restored. $M$ is the constituent quark
mass. The dashed and the solid lines correspond respectively to the chiral crossover and to the chiral first order
transition, the dot denotes the CEP. In the gray region chiral symmetry is restored but a bound state with the quantum
numbers of the pions can still be formed. See the text for more details.}
\end{figure}

It is interesting to evaluate the ratio $m_\pi/2M$, with $M$ the constituent quark mass, as a function of temperature.
If it is larger than one, than a bound state with the quantum numbers of the pion is less stable than a state made of a
free quark and a free antiquark, and thus the pion melts to its constituent quarks. In Fig.~\ref{FIG:ccc} where we show
the region in the $\mu-T$ plane where the bound state can be formed. In the white region below the gray domain the
chiral symmetry is broken and pions live as pseudo-Goldstone modes. In the gray region chiral symmetry is restored but
$m_\pi/2M_q < 1$, thus bound states can be formed. Finally, in the white region above the gray domain chiral symmetry
is restored and free quark states are more stable than bound states (this region is characterized by $m_\pi/2M_q > 1$).
The window in which the bound state is stable above the chiral crossover is not a peculiarity of the PNJL model.
Similar results are obtained in Refs.~\cite{Hatsuda:1985eb,Hansen:2006ee}

In conclusion, we have studied the phase diagram of the two flavor PNJL model, considering the role of electrical
neutrality and finite quark masses on the chiral crossover and the pion condensation. One of our results is that pion
condensation does not occur in the electrical neutral state.

We have computed the masses of the pseudo-Goldstone modes and of the $\sigma$-mode. Furthermore we have investigated on
the possibility of existence of bound states with the quantum numbers of the pions above the chiral critical
temperature.  The result is summarized in Fig.~\ref{FIG:ccc}.

\section*{Acknowledgements}
I would like to thank H. Abuki, R. Gatto, N. Ippolito and G. Nardulli for the pleasant collaboration. Moreover I
aknowledge  M. Hamada, M. Huang, O. Kiriyama,  V. A. Miransky, K. Redlich, C. Sasaki, A. Schmitt, I. Shovkovy, W. Weise
for discussions made during the Workshop ``New Frontiers in QCD 2008''. Finally I would like to thank K. Fukushima and
T. Kunihiro for their kind invitation to the aforementioned Workshop.


\end{document}